\documentclass[letterpaper]{JHEP3}

\usepackage{epsfig}

\def\endignore{}
\def\ignore #1\endignore{} 

\def\Box{{\hbox{$\sqcup$}\llap{\hbox{$\sqcap$}}}}

\def \gsim{\mathrel{\vcenter
     {\hbox{$>$}\nointerlineskip\hbox{$\sim$}}}}

\def\be{\begin{equation}}
\def\ee{\end{equation}}
\def\bea{\begin{eqnarray}}
\def\eea{\end{eqnarray}}
\def\nn{\nonumber}
\def\ansatz{{\it ansatz}}

\def\exd{{\rm d}}
\def\pref#1{(\ref{#1})}
\def\endignore{}
\def\ignore #1\endignore{} 

\def\bd{\begin{displaymath}}
\def\ed{\end{diplaymath}}

\def\d{\mathrm{d}}

\def\bfE{{\bf E}}

\def\bfr{{\bf r}}

\def\cW{{\cal W}}
\def\cA{{\cal A}}

\def\ssM{{\scriptscriptstyle M}}
\def\ssN{{\scriptscriptstyle N}}
\def\ssP{{\scriptscriptstyle P}}

\def\ba{\begin{eqnarray}}
\def\ea{\end{eqnarray}}
\def\be{\begin{equation}}
\def\ee{\end{equation}}

\def\exd{{\rm d}}

\def\({\left(}
\def\){\right)}
\newcommand{\eqref}[1]{(\ref{#1})}

\title{Scaling Solutions to 6D Gauged Chiral Supergravity}

\author{A. J. Tolley,${}^1$ C.P. Burgess,${}^{2,3}$ C. de Rham${}^4$ and D. Hoover${}^4$\\
${}^1$ Joseph Henry
Laboratories, Princeton University, Princeton NJ, 08544, USA.\\
${}^2$ Dept. of Physics \& Astronomy, McMaster University,
Hamilton ON, L8S 4M1, Canada. \\
${}^3$ Perimeter Institute for Theoretical Physics, Waterloo ON, Canada. \\
${}^4$ Physics Department, McGill University, Montr\'eal QC, H3A
2T8, Canada. \\
}

\date{}

\abstract {We construct explicitly time-dependent exact solutions
to the field equations of 6D gauged chiral supergravity,
compactified to 4D in the presence of up to two 3-branes situated
within the extra dimensions. The solutions we find are
scaling solutions, and are plausibly attractors which represent
the late-time evolution of a broad class of initial conditions. By
matching their near-brane boundary conditions to physical brane
properties we argue that these solutions (together with the known
maximally-symmetric solutions and a new class of
non-Lorentz-invariant static solutions, which we also present
here) describe the bulk geometry between a
pair of 3-branes with non-trivial on-brane equations of state. }

\begin{document}

\section{Introduction}

The discovery of the existence of D-branes \cite{DBranes} within
string theory has led to a fundamental rethinking of the kinds of
effective theories which can describe the low-energy limit of a
fundamental theory. The broadening of the collective mind to which
this rethinking has led has permitted new progress to be made on a
number of the `naturalness' issues which seem to plague our
understanding of the low-energy theories --- {\it i.e.} the
Standard Model plus General Relativity --- which describe the
world we see around us.

This progress has come about largely because the possibility of trapping
low-energy particles on surfaces within higher-dimensional
spacetime changes how one must think about naturalness issues. For
instance, an important property of 4 dimensions is the equivalence
of a vacuum energy with a cosmological constant, and so also with
nonzero 4D curvature. This connection underlies the cosmological
constant problem \cite{CCproblem1,CCproblem2}, which amounts to
the difficulty in understanding why the observed universe can be
so flat given that quantum fluctuations generically make the
vacuum energy enormously large.

This connection can be broken in higher-dimensional brane
configurations, inasmuch as known higher-dimensional solutions
show that large 4D energy sources can coexist with zero 4D
curvature, as has been pointed out for pure gravity with
co-dimension one \cite{5DSelfTune,5DSelfTunex} and co-dimension
two \cite{6DNonSUSYSelfTune,6DNonSUSYSelfTunex,NavSant}
geometries, as well as for co-dimension two geometries within
supergravity \cite{SLED1,SLED2,SLEDx,GGPplus}. They can coexist in
this way because within the extra-dimensional context the 4D
energy density turns out to source the curvature of the extra
dimensions rather than the curvature of the 4 dimensions within
which the energy density exists.

Whether or not this observation can lead to a solution of the
cosmological constant problem depends on identifying the
assumptions which are required in order to ensure that the 4D
curvature is sufficiently small, and on whether or not these
choices are natural or if they must be fine tuned. The hope is
that the reformulation of the cosmological constant problem in
this way can usefully recast the issues into a form which might
allow progress that was not possible in the 4D context. 6D
supergravity provides a particularly attractive framework within
which to test these tuning issues in detail, since in six
dimensions it is possible to have internal geometries which are
large enough to allow the Casimir energy in these dimensions to be
of order the size of the observed cosmological constant, allowing
a variety of attractive phenomenological possibilities
\cite{SLEDpheno}. In this paper we use 6D chiral gauged
supergravity to explore one of the fine-tuning issues which arise
within this extra-dimensional context.

There are two different kinds of fine-tunings against which one
must be vigilant within these extra-dimensional models. The first
of these is the requirement to tune against quantum fluctuations.
That is, if parameters are chosen to ensure that 4 dimensions are
flat, do these dimensions remain flat once the scale of the theory
is lowered by integrating out particles having successively lower
masses. This is the traditional cosmological constant problem, and
this paper has nothing to add to the ongoing investigations as to
whether the higher-dimensional models require tuning in this way
\cite{UVSensitivity}. Our focus here is instead on the second
fine-tuning issue, which arises in any formulation (such as the
higher-dimensional scenarios of interest) for which the Dark
Energy density is time-dependent. This second question asks
whether having an acceptable present-day cosmology requires an
inordinately finely tuned adjustment of initial conditions before
the advent of the present epoch's Big Bang cosmology.

In order to do this we shall construct several nontrivial exact
solutions of the Nishino-Sezgin 6D chiral gauged supergravity
\cite{NS,6DSugra,HiDSugra}.\footnote{See also ref.~\cite{ED} for a
class of solutions which are very similar to some of the ones we
present here.} These solutions will correspond, in general, to the
introduction of non-trivial matter sources on the brane, but in
specific cases they describe the response of the bulk geometry to
pure tension branes whose tensions are not otherwise fine-tuned.
In this sense we are probing the cosmological constant problem in
these six dimensional compactifications, by inferring the brane
geometries induced, via the bulk geometry, from arbitrary brane
tensions. From a cosmological point of view it is also interesting
to have explicit solutions for branes with arbitrary equations of
state, and couplings to the bulk fields such as the dilaton. We
take a pragmatic approach of first looking for bulk solutions, and
then inferring the properties of the branes for which these would
be consistent solutions.

The remainder of the paper is organized as follows: we close this
section with a brief summary of the relevant field equations. \S2\
then describes the properties of a 3-parameter class of static
solutions having two compact dimensions. They include the
2-parameter family of previously-known solutions which are
maximally symmetric in the four noncompact dimensions, as well as
a new set of static solutions which break the 4D Lorentz symmetry.
These solutions generically contain singularities which we can associate with
the positions of two source branes. \S3\ derives a broad class of
time-dependent solutions describing geometries for which the sizes
of the various dimensions scale with a power of time. A related
class of solutions describing one brane within two noncompact
dimensions is then described in \S4. In \S5\ we argue that these
solutions are rich enough to describe the late-time behaviour of
a large class of source branes. We provide the
matching conditions which allow the asymptotic behaviour of the
bulk fields to be related to physical properties of these branes
like energy density, pressure and dilaton coupling. We close in
\S6\ with a brief summary and conclusions.

\subsection*{Field equations}

The bosonic part of the Lagrangian density for 6D chiral gauged
supergravity \cite{NS,6DSugra,HiDSugra} is given by\footnote{The
curvature conventions used here are those of Weinberg's book
\cite{GandC}, and differ from those of MTW \cite{MTW} only by an
overall sign in the Riemann tensor.}
\be
    \frac{{\cal L}}{\sqrt{-g}} = - \frac{1}{2 \kappa^2} \,
    g^{\ssM\ssN} \Bigl[ R_{\ssM\ssN} + \partial_\ssM \phi \,
    \partial_\ssN \phi \Bigr] - \frac14 \, e^{-\phi} \, F_{\ssM\ssN}
    F^{\ssM\ssN} - \frac{2g^2}{\kappa^4} \; e^\phi  \,,
\ee
where $\phi$ is the 6D scalar dilaton and $F = \exd A$ is the
appropriate field strength for the gauge potential, $A_\ssM$,
which gauges a specific abelian $R$-symmetry, whose gauge
coupling, $g$, has dimensions of inverse mass. We keep the 6D
Planck scale, $\kappa^2 = M_6^{-4}$, explicit for ease of
comparison with the various conventions which are used in the
literature. These expressions set to zero some of the bosonic
fields of 6D supergravity, including possible matter hyperscalars
and gauge potentials, $\Phi^a = A^\alpha_\ssM = 0$, and
Kalb-Ramond fields, $G_{\ssM\ssN\ssP} = 0$
--- a choice which is consistent with the corresponding field
equations \cite{HypersNonzero}.

This action leads to the following field equations:
\bea \label{fieldequations}
    D_\ssM \Bigl(e^{ - \phi} \, F^{\ssM\ssN} \Bigr) &=& 0 \qquad
    \hbox{(Maxwell)} \nn\\
    \Box \, \phi + \frac{\kappa^2}{4} \, e^{-\phi} \;
    F_{\ssM\ssN} F^{\ssM\ssN} - \frac{2 \, g^2}{\kappa^2} \, e^{\phi}
    &=& 0 \qquad \hbox{(Dilaton)} \nn\\
    R_{\ssM\ssN} + \partial_\ssM\phi \, \partial_\ssN\phi
    + \kappa^2 e^{- \phi} \; F_{\ssM\ssP} {F_\ssN}^\ssP
    + \frac12 \, (\Box\,
    \phi)\, g_{\ssM\ssN} &=& 0 \qquad \hbox{(Einstein)}\,,\nn\\
\eea
to which the bulk of the remainder of this paper is dedicated to
solving, in order to find explicit static and time-dependent
compactifications to four dimensions. To this end we divide the
six coordinates $x^\ssM$, $M = 0,\dots,5$, into 4D coordinates
$x^\mu$, $\mu = 0,\dots,3$, and 2D coordinates $x^m$, $m = 4,5$.
When required, the three spatial coordinates of the noncompact 4
dimensions are denoted $x^i, i = 1,2,3$.

\section{Static solutions}
\label{static}

We start by describing a broad class of static compactifications.
For some of these solutions the noncompact 4 dimensions are
maximally symmetric, and these solutions have been described in
the literature. We supplement these with new static solutions
which break the 4D Lorentz symmetry.

\subsection{Ans\"atze}

\label{staticsolutionssection}

We start by writing out the field equations for configurations
which are ($i$) time-independent; ($ii$) translation and rotation
invariant in the 3 noncompact spatial dimensions; and ($iii$) are
axially symmetric in the extra dimensions. That is, we take
\bea
    \exd s^2 &=& -e^{2\hat w(\eta)} \, \exd t^2 + e^{2\hat a(\eta)} \,
    \delta_{ij} \, \exd x^i \exd x^j
    + e^{2 \hat v(\eta)} \, \exd \eta^2 + e^{2 \hat b(\eta)}
    \exd \theta^2  \nn\\
    A_{\theta} &=& \hat a_\theta(\eta) \qquad \hbox{and} \qquad
    e^{\phi} = e^{\hat\varphi(\eta)} \,,
\eea
leading to the following system of coupled ordinary differential
equations:
\bea \label{BackgroundMaxwell}
    \hat a_\theta'' + \Bigl( \hat w' + 3\, \hat a'  - \hat v' - \hat b' -
    \hat\varphi' \Bigr) \, \hat a_\theta'
    &=& 0 \qquad \hbox{(Maxwell)} \nn\\
    \hat\varphi'' + \Bigl(\hat w' + 3\,\hat a' -\hat v'
    + \hat b' \Bigr) \hat\varphi'
    + \frac{\kappa^2}{2}\, e^{-2 \hat b -\hat\varphi} (\hat a_\theta')^2
    - \frac{2 g^2}{\kappa^2} \, e^{2\hat v + \hat\varphi} &=& 0
    \qquad \hbox{(Dilaton)} \nn\\
    \hat w'' + \Bigl(\hat w' + 3\,\hat a' - \hat v' + \hat b') \hat w'
    - \frac{\kappa^2}{4} \, e^{- 2\hat b - \hat\varphi} (\hat a_\theta')^2
    + \frac{g^2}{\kappa^2} \, e^{2\hat v + \hat\varphi}
    &=& 0 \qquad \hbox{($tt$ Einstein)} \nn\\
    \hat a'' + \Bigl(\hat w' + 3\,\hat a' - \hat v' + \hat b') \hat a'
    - \frac{\kappa^2}{4} \, e^{- 2\hat b - \hat\varphi} (\hat a_\theta')^2
    + \frac{g^2}{\kappa^2} \, e^{2\hat v + \hat\varphi}
    &=& 0 \qquad \hbox{($ii$ Einstein)} \nn\\
    \hat b'' +  \Bigl(\hat w' + 3\,\hat a' - \hat v' + \hat b' \Bigr) \hat b'
    + \frac{3\kappa^2}{4} \,
    e^{- 2\hat b -\hat\varphi} (\hat a_\theta')^2
    + \frac{g^2}{\kappa^2} \, e^{2\hat v + \hat\varphi}
    &=& 0 \qquad \hbox{($\theta\theta$ Einstein)} \nn\\
    \hat w'' + 3\, \hat a'' + \hat b'' + (\hat w')^2 + 3\, (\hat
    a')^2 + (\hat b')^2 + (\hat\varphi')^2
    \qquad\qquad && \phantom{=} \qquad \hbox{($\eta\eta$ Einstein)} \nn\\
    - \Bigl(\hat w' + 3\,\hat a' + \hat b' \Bigr) \hat v'
    + \frac{3\kappa^2}{4} \,
    e^{- 2\hat b - \hat\varphi} (\hat a_\theta')^2
    + \frac{g^2}{\kappa^2}\, e^{2 \hat v + \hat\varphi} &=& 0 \,.
\eea
Here $'$ denotes a derivative with respect to $\eta$. These equations must be supplemented with boundary conditions at the locations of the two branes.

Although this appears to provide 6 equations for the 6 unknown
functions $\hat w$, $\hat v$, $\hat b$, $\hat\varphi$ and $\hat
a_\theta$, this is deceptive because we can ensure that one of
these functions (say $\hat v$) takes any particular form simply by
appropriately changing the coordinate $\eta$. However, one
combination of these 5 equations, found by taking the combination
$(\eta\eta) - (tt) - 3\, (ii) - (\theta\theta)$, can be thought of
as a `constraint' on the evolution of the fields into the $\eta$
direction because all second derivatives, $(d/d\eta)^2$, drop
out\footnote{This equation follows directly from writing the
$(\eta\eta)$ Einstein equation in terms of the Einstein tensor,
$G_{\eta\eta} + \kappa^2 T_{\eta\eta} = 0$, and it is the Bianchi
identity which ensures that it holds for all $\eta$ once it is
imposed on `initial conditions' at $\eta = \eta_0$. This is the
Hamiltonian constraint associated with the lapse function
$\exp{\hat{v}}$ in the spacelike ADM formalism of gravity.}:
\be \label{Constraint}
    \kappa^2 \, e^{- 2\hat b - \hat\varphi} (\hat a_\theta')^2
    =  6\,(\hat a')^2  + 6\, \hat a' \hat w' + 6\, \hat a' \hat b'
    + 2\, \hat b' \hat w'  - (\hat\varphi')^2
    + \frac{4g^2}{\kappa^2} \, e^{2\hat v + \hat\varphi} \,.
\ee
As a consequence of the Bianchi identities, one of the
remaining 5 equations is then not independent and can be derived
from the derivative of the constraint and the other 4 equations.

A considerable simplification to this system of equations can be
obtained by performing the redefinitions
\ba
    \hat{w} &=& 3 \xi+(y-x)/4 \nn \\
    \hat{a} &=& -\xi +(y-x)/4 \nn \\
    \hat{v} &=& \ln N + (5y-x+2z)/4 \nn \\
    \hat{b} &=& (3 x+y+2 z)/4 \nn \\
    \hat{\varphi} &=& (x-y-2z)/2.
\ea
Then one can demonstrate that the system of equations
(\ref{BackgroundMaxwell}) and the constraint (\ref{Constraint})
follow from the Euler-Lagrange equations associated with the
action
\be
    S=\int \d \eta \, \left[ N^{-1}
    \left[(x')^2-(y')^2+(z')^2+12 (\xi')^2+\kappa^2
    e^{-2x}(\hat{a}_{\theta}')^2 \right] +N \left[
    \frac{4g^2}{\kappa^2}e^{2y}\right] \right] \,,
\ee
here $N$ plays the role of a Lagrange multiplier and can be set to
unity after variation. This system can be solved exactly, giving
(with $N=1$):
\ba
    \hat{a}_{\theta} &=& q \int \d \eta \, e^{2x} \nn \\
    \nn e^{-x} &=&
    \frac{\kappa q }{\lambda_1} \cosh[\lambda_1 \, (\eta-\eta_1) ]  \\
    \nn e^{-y} &=&   \frac{2 g}{\kappa \lambda_2}\cosh[\lambda_2 \,
    (\eta-\eta_2) ] \\ \nn
    z &=& z_0+ \lambda_3 \, \eta \\
    \xi &=& \xi_0 + \lambda_4 \, \eta
\ea
and where the constraint equation amounts to the condition
$\lambda_2^2=\lambda_1^2+\lambda_3^2+12 \,\lambda_4^2$. We can set two of the parameters $\eta_1,\eta_2,z_0,\xi_0$ to zero by coordinate rescalings without loss of generality.

\subsection{Lorentz-invariant solutions}

A special case of these solutions is $\lambda_4=0$, for which
$\xi'=0$ and so $\hat{a}'=\hat{w}'$. The resulting solutions are
therefore Lorentz invariant, and reduce to those that have been
found earlier in the literature
\cite{GGP,SLED2,GGPplus,6DdSSUSY}.\footnote{The conventions of
ref.~\cite{GGP} may be obtained from ours by taking $R_{MN} \to -
R_{MN}$, $\phi \to -\phi/2$ and $\kappa^2 = 1/2$, while those of
\cite{SLED2} differ from those here only by the choice $\kappa^2 =
1$.} In this case the general solutions can be written as
\bea \label{GGPFields}
    \exd s^2 &=& \cW^2(\eta) \, \eta_{\mu\nu} \, \exd x^\mu \exd x^\nu
    + \cA^2 (\eta) \Bigl[ \cW^8(\eta) \, \exd \eta^2 +
    \exd \theta^2 \Bigr] \nn\\
    F_{\eta\theta} &=& \left( \frac{q \cA^2 }{\cW^2} \right)
    e^{-\lambda_3 \eta} \qquad \hbox{and} \qquad
    e^{-\phi} = \cW^2 \, e^{\lambda_3 \eta} \,,
\eea
where
\bea \label{GGPSolutions}
    \cW^4 &=& \left( \frac{\kappa^2 q \lambda_2}{2 g \lambda_1}
    \right) \frac{\cosh[\lambda_1(\eta -
    \eta_1)]}{\cosh[\lambda_2(\eta-\eta_2)]} \nn\\
    \cA^{-4} &=& \left( \frac{2 \kappa^2 q^3 g}{\lambda_1^3
    \lambda_2} \right) e^{-2\lambda_3 \eta} \cosh^3[
    \lambda_1(\eta -\eta_1)] \cosh[ \lambda_2(\eta - \eta_2)] \,,
\eea
along with the constraint $\lambda_2^2 = \lambda_1^2 +
\lambda_3^2$.

A further special case is obtained for $\lambda_3=\lambda_4=0$,
and $\eta_1=\eta_2$ for which $\hat \varphi = \hat\varphi_0$ and
$\hat w = \hat a = \hat w_0$ are constants. Changing variables to
proper distance, $\exd \rho = e^{\hat v(\eta)} \exd \eta$, the
form field becomes
\be
    F_{\rho\theta} = \hat a_\theta'
    = \pm \frac{2g}{\kappa^2} \, e^{\hat\varphi_0} \, \hat B(\rho) \,,
\ee
where primes now indicate differentiation with respect to $\rho$
and $\hat B(\rho) \equiv e^{\hat b(\rho)}$ satisfies
\be
    \frac{\hat B''}{\hat B} = \hat b'' + (\hat b')^2
    = - \frac{4g^2}{\kappa^2} \, e^{\hat\varphi_0} \,,
\ee
with solution $\hat B(\rho) = B_0 \sin\Bigl(2g e^{\hat\varphi_0/2}
\rho/\kappa\Bigr)$. These are the rugby-ball generalizations
\cite{SLED1} of the older Salam-Sezgin solution \cite{NS} to
gauged chiral 6D supergravity.

\subsection{Asymptotic Forms}
\label{asymptoticforms}

These solutions describe geometries which become singular at $\eta
= \pm \infty$, which are interpreted as being the positions of the
3-branes which source this configuration. Since it is ultimately
the internal structure of the brane (if any) which is responsible
for resolving these singularities, one expects this structure to
be related to the asymptotic limit of the above solutions in the
near-brane limit. We therefore pause here to outline what this
asymptotic near-brane behaviour is.

To this end it is useful to adopt Gaussian-normal (GN) coordinates
near the brane for which $\exd s^2 = \hat g_{ab} \, \exd x^a \exd
x^b + \exd \rho^2$,  where $x^a$ denotes the 5 other coordinates,
$\{ x^a \} = \{ x^\mu, \theta \} = \{ t,x^i,\theta \}$, with the
brane position being described by $\rho = 0$. We take the
asymptotic form of the bulk fields in the near-brane limit ($\rho
\gsim \ell$) to be generically given by a power law
\cite{6DdSSUSY}
\bea \label{asymptoticfields}
    &&\exd s^2 \sim - [c_w (H_1\rho)^{\omega}]^2 \, \exd t^2
    + [c_a (H_1\rho)^{\alpha}]^2\, \delta_{ij} \, \exd x^i \exd x^j
    +  \exd \rho^2 + [c_\theta (H_1\rho)^{\beta-1}]^2 \rho^2 \exd \theta^2
    \nn\\
     &&\qquad\qquad e^{\phi} \sim c_\phi (H_1\rho)^p
        \quad \hbox{and} \quad
        F^{\rho\theta} \sim c_f \, (H_1\rho)^\gamma\,,
\eea
where $\omega$, $\alpha$, $\beta$, $p$, $\gamma$, $c_w$, $c_a$,
$c_\theta$, $c_\phi$ and $c_f$ are constants, and $H_1$ is an
arbitrary scale. With these choices the extrinsic curvature,
$K_{ab} = \frac12 \, \partial_\rho \hat g_{ab}$, of the
constant-$\rho$ surfaces becomes
\be
    {K^t}_t \approx \frac{\omega}{\rho} \,, \quad
    {K^i}_j \approx \frac{\alpha}{\rho} \, {\delta^i}_j \quad
    \hbox{and} \quad {K^\theta}_\theta \approx \frac{\beta}{\rho}
    \,,
\ee
up to contributions that are subleading for small $\rho$.

Only two of the five powers $\alpha$, $\beta$, $\omega$, $\gamma$
and $p$ defined above are independent, since the bulk field
equations impose the following two conditions amongst them
\be \label{PowerFieldEqns}
    \omega^2 + 3 \alpha^2 + \beta^2 + p^2 = \omega + 3 \alpha +
    \beta = 1 \quad \hbox{and} \quad
    \gamma = p -1 \,.
\ee
For example, explicit calculation with the general static
solutions given above gives
\ba \label{GGPPowers}
    && \alpha_\pm = \frac{\lambda_2 - \lambda_1 \pm 4
    \lambda_4}{5 \lambda_2 - \lambda_1
    \mp 2 \lambda_3} \,,  \quad \quad \quad \omega_\pm =
    \frac{\lambda_2 - \lambda_1 \mp 12 \lambda_4}{5 \lambda_2 - \lambda_1
    \mp 2 \lambda_3}\, , \\
    && \beta_\pm = \frac{\lambda_2 + 3 \lambda_1 \mp 2\lambda_3}{5
    \lambda_2 - \lambda_1 \mp 2 \lambda_3}\, , \quad
    \hbox{and} \quad
    p_\pm = -\frac{2( \lambda_2 - \lambda_1 \mp 2 \lambda_3)}{5
    \lambda_2 - \lambda_1
    \mp 2 \lambda_3}  \,,
\ea
at the brane positions $\eta \to \pm \infty$. As is easily
verified, these satisfy the expressions \pref{PowerFieldEqns} above.

In later sections we relate these powers to the physical
properties of the branes which source these geometries, following
arguments which generalize those of
refs.~\cite{GGPplus,NavSant,6DdSSUSY}.

\section{Scaling solutions}

We now generalize the previous discussion to a new class of
time-dependent scaling configurations which provide exact
solutions to the same 6D field equations. The idea behind the
construction is to assume a metric of the general form
\be \label{MetricScalingAnsatz0}
    \exd s^2 = - t^c N^2 \exd t^2
    +q_{ij}\(t^{c/2}\beta^i \exd t+t^{c/2+1}\exd x^i\)\(t^{c/2}\beta^j \exd t+t^{c/2+1}\exd
    x^j\),
\ee
for arbitrary real $c$, where $N$ and $\beta^i$ represent the lapse and shift
functions just as in the usual ADM decomposition and $q_{ij}$ is the
metric in the $n$-dimensional hypersurface with normal
vector $\beta^i$. These quantities are
assumed to be independent of time: $N=N(x^i)$, $\beta^j=\beta^j(x^i)$ and
$q_{ij}=q_{ij}(x^i)$. This is the most general metric, up to coordinate transformations, preserving a single time-like homothetic Killing vector that acts as $t \rightarrow \lambda t$, $ g_{MN} \rightarrow \lambda^{2+c}g_{MN}$.

The key point to recognize is that with this choice the
components of the Ricci tensor also scale as a simple power of
$t$: $R_{tt} \propto t^{-2}$, $R_{t i} \propto t^{-1}$ and
$R_{ij} \propto t^{0}$. Because of this, and of the scale-invariance
of the supergravity equations, it is also possible to scale the
other fields in the problem with $t$ in such a way as to ensure
that the field equations are also proportional to specific powers
of $t$. Once this is done, then the problem of solving the field
equations becomes an exercise in determining the profiles of
the metric functions in one lower dimension.

The most general form for the dilaton and the gauge field necessary for eq. (\ref{MetricScalingAnsatz0}) to be a consistent solution of the equations of motion is determined by the scaling symmetry to be
\bea
\phi &=& \bar{\phi}(x^i)-(2+c)\log t\\
A_\mu &=&\(\begin{array}{r}
t^{-1}A_0(x^j)\\ A_i(x^j)
\end{array}\).
\eea
On substitution of these forms into the 6d equations of motion we find that the time dependence completely decouples as expected
from the scaling symmetry, and we are left with an effective 5 dimensional theory for the profiles $q_{ij},\beta_i,N,\bar{\phi},A_0,A_i$. In practice the resulting 5d system is formidable and so in what follows we shall consider
special cases of these scaling solutions. In particular we
concentrate on the case where the variables only depend on one of the spatial
dimensions (namely the radial direction $\eta$). Furthermore we will
work in the special case where only one component of the gauge field
does not vanish, in particular $A_0=0$. All these assumptions
greatly simplify this effective theory.

\subsection{Warped scaling solutions}

\label{constraintssolutions}

From now on we shall only consider scaling solutions which are warped on a single extra dimension.
We consider the special case
where the metric is
\be \label{MetricScalingAnsatz}
    \exd s^2 = (H_0 t)^c  \Bigl[ - e^{2w(\eta)}
    \exd t^2 + e^{2a(\eta)} \, \delta_{ij} \, \exd x^i \exd x^j
    \Bigr] + (H_0 t)^{2+c} \Bigl[ e^{2v(\eta)}  \exd \eta^2 +
    e^{2b(\eta)} \exd \theta^2 \Bigr]  \,,
\ee
while
\be
    A_\theta = a_\theta(\eta) \qquad \hbox{and} \qquad
    e^\phi = \frac{e^{\varphi(\eta)}}{(H_0 t)^{2+c}}  \,.
\ee
Although this metric is not precisely of the form of eq. (\ref{MetricScalingAnsatz0}) it can easily be shown to be equivalent by redefining $x^i=t \tilde{x}^i$. From now on we shall work with the form (\ref{MetricScalingAnsatz}) for convenience.
Here $H_0$ is a constant of dimension inverse time and the
functions $w$, $a$, $v$, $b$, $a_\theta$ and $\varphi$ are to be
determined by solving the field equations. In some cases it will be natural to take $H_0<0$ so that the direction of increasing time corresponds to $t \rightarrow 0^{-}$.

By virtue of the way the Ricci tensor scales with $t$, with the
above ansatz all of the field equations reduce to the following
set of coupled ordinary differential equations which govern the
$\eta$-dependence of the various undetermined functions. The
Maxwell equation is
\be
    a_\theta'' + \Bigl( w' + 3\, a' - b' -  v' - \varphi'
    \Bigr) a_\theta'
    = 0  \,,
\ee
while the Dilaton equation similarly becomes
\bea
    \varphi'' + \Bigl(w'+ 3\, a' - v' + b' \Bigr) \varphi'
    + (2+c)(2c+1) \, e^{2(v-w)} H_0^2
    \qquad\qquad && \\
    + \frac{\kappa^2}{2} e^{-2b - \varphi} (a_\theta')^2 -
    \frac{2g^2}{\kappa^2} \, e^{2v + \varphi}
    &=& 0 \,. \nn
\eea
The Ricci tensor for this class of metrics is easily computed and
leads to the following components for the Einstein equations. The
$(t\eta)$ component is
\be \label{tetaConstraint}
    (2c+1)\, w' +3\,a' +(2+c)\, \varphi' =0 \,,
\ee
while the $(tt)$ equation is
\be
    w''
    + \Bigl( w' + 3a' - v' + b' \Bigr) w'
    + \frac{\kappa^2}{4} e^{-2b - \varphi} (a_\theta')^2
    + \frac{g^2}{\kappa^2} \, e^{2v + \varphi} 
    - \left( c^2 + \frac52 \, c + 4 \right)
    e^{-2w + 2v} H_0^2 = 0 \,.
\ee
The $(\theta\theta)$ equation is
\be
    b'' + \Bigl( w' + 3a' - v' + b' \Bigr) b' + \frac{3\kappa^2}{4}
    e^{-2b - \varphi} (a_\theta')^2 + \frac{g^2}{\kappa^2} \,
    e^{2v + \varphi} 
    - \frac12 (c+2) (2c+1)
    e^{-2w + 2v} H_0^2=0 \,,
\ee
the $(ii)$ Einstein equation becomes
\be
    a''
    + \Bigl( w' + 3a' - v' + b' \Bigr) a'
    - \frac{\kappa^2}{4} e^{-2b - \varphi} (a_\theta')^2
    + \frac{g^2}{\kappa^2} \, e^{2v + \varphi} 
    - \frac12 c (2c+1)
    e^{-2w + 2v} H_0^2 = 0 \,,
\ee
and finally the $(\eta\eta)$ equation is
\bea
    &&w'' + 3a'' + b'' + (w')^2 + 3(a')^2 + (b')^2 + (\varphi')^2
    - \Bigl( w' + 3a' + b' \Bigr) v' \\
    &&\qquad\qquad\qquad\qquad
    + \frac{3\kappa^2}{4} e^{-2b - \varphi} (a_\theta')^2
    + \frac{g^2}{\kappa^2} \, e^{2v + \varphi}
    - \frac12 (c+2) (2c+1)
    e^{-2w + 2v} H_0^2 = 0 \,, \nn
\eea
where $'=d/d\eta$.
Equation counting proceeds much as for the static solutions,
although with an important difference. The difference is the
inclusion of time dependence, which makes two of the components of
the Bianchi identity nontrivial rather than one. This implies that
in this case {\it two} of the field equations are not independent
of the others, rather than just one. Related to this is the
existence in this case of {\it two} constraint equations which do
not involve $\exd^2/\exd \eta^2$, which the Bianchi identities
ensure are preserved when evolved using the field equations in the
$\eta$ direction. These constraints can be taken to be the
$(t\eta)$ Einstein equation, eq.~\pref{tetaConstraint}, and the
combination $(\eta\eta) - (tt) - 3(ii) - (\theta\theta)$:
\be \label{diagConstraint}
    (\varphi')^2
    - 6 ( w' + a' + b') a' - 2 b' w'
    + {\kappa^2} e^{-2b - \varphi} (a_\theta')^2
    - \frac{4g^2}{\kappa^2} \, e^{2v + \varphi}
    + 4 (c^2 + c+1)
    e^{-2w + 2v} H_0^2 = 0 \,. \nn
\ee
The logic to solving these equations is to use
eq.~\pref{tetaConstraint} to solve for $a$ and
eq.~\pref{diagConstraint} to solve for $a_\theta$. Once this is
done, we may ignore the Maxwell and $(ii)$ Einstein equations as
they are redundant. We are then left with three independent equations
--- {\it i.e.} the Dilaton and the $(tt)$ and $(\theta\theta)$
Einstein equations --- for evolving the remaining three
independent functions --- $w$, $b$ and $\varphi$ --- into the
$\eta$ direction.

As for the static case the system of equations is simplified by
making the choice
\ba\label{redefinition}
    w &=& 3 \xi+(y-x)/4 \nn \\
    a &=& -\xi +(y-x)/4 \nn \\
    v &=& \ln N + (5y-x+2z)/4 \nn \\
    b &=& (3 x+y+2 z)/4 \nn \\
    \varphi &=& (x-y-2z)/2.
\ea
The $(t \eta)$ equation now amounts to the condition
\be
    \xi(\eta) \equiv \frac14[w(\eta) - a(\eta)] =\frac{(2+c)}{6 c} \, z(\eta) \,,
\ee
up to an irrelevant integration constant. The remaining equations
of motion and $(\eta \eta)- (tt) - 3(ii) - (\theta\theta)$ follow
from the Euler-Lagrange equations of the action
\ba
    S &=& \int \d \eta \left\{ N^{-1} \left[(x')^2-(y')^2
    +\frac{4(1+c+c^2)}{3c^2}(z')^2
    +\kappa^2 e^{-2x}(a_{\theta}')^2 \right] \right. \nn \\
    &&\qquad\qquad\qquad\qquad + \left.  N \left[ \frac{4g^2}{\kappa^2}
    \, e^{2y}-4 H_0^2 (1+c+c^2)e^{2y-2 z/c}
    \right]  \right\}\,,
\ea
where $N$ plays the role of a Lagrange multiplier and can be set
to unity after variation. As before $a_{\theta}$ and $x$ decouple
and can be explicitly integrated:
\ba
    a_{\theta} &=& q \int \d \eta \, e^{2x} \nn \\
    e^{-x} &=& \frac{\kappa q }{\lambda_1} \cosh[\lambda_1 \, (\eta-\eta_1) ]
    \,,
\ea
but $y$ and $z$ do not. Their equations follow from the reduced
action
\ba
    S &=& \int \d \eta \left\{
    N^{-1} \left[-(y')^2+\frac{4(1+c+c^2)}{3c^2}(z')^2 \right]
    \right.\nn \\
    && \qquad\qquad
    + \left. N \left[ -\lambda_1^2 +\frac{4g^2}{\kappa^2}e^{2y}-4 H_0^2
    (1+c+c^2)e^{2y-2 z/c}\right] \right\} .
\ea
With the choice $N=1$ the resulting field equations are
\ba
    y''+\frac{4g^2}{\kappa^2} e^{2y}-4 H_0^2 (1+c+c^2)e^{2y-2 z/c} &=& 0 \\
    z''-3 c H_0^2 e^{2y-2 z/c} &=& 0.
\ea
The asymptotic form of these solutions as $\eta \rightarrow \pm
\infty$ is $y \rightarrow \lambda_2^{\pm}\, \eta$ and $z
\rightarrow \lambda_3^{\pm}\, \eta$ and the constraint implies
$(\lambda_2^{\pm})^2=\lambda_1^2+4(1+c+c^2)(\lambda_3^{\pm})^2/(3c^2)$.
Although we have not found closed-form analytic solutions of these
equations, they are straightforward to integrate numerically, with
the result having qualitatively the same form as the de Sitter
solutions considered in \cite{6DdSSUSY}.\footnote{See also
\cite{6DdSNonSUSY} for de Sitter solutions in the
non-supersymmetric case.} This relation with the de Sitter
solutions is not surprising since these solutions may be obtained
as the special case $c=-2$ of the above scaling solutions.

\subsection{Useful special cases}

There are several special cases of the previous solutions which
are of particular interest.

\subsubsection*{Connection to 4D scaling solutions}

If $c = -1$ then the metric and dilaton have the time dependence
\be \label{MetricScalingAnsatz00}
    \exd s^2 = \frac{1}{t} \, \hat g_{\mu\nu}(y) \, \exd x^\mu \exd x^\nu
    + t \, \hat g_{mn}(y) \, \exd y^m \exd y^n
    \quad\hbox{and}\quad
    e^{\phi} = \frac{e^{\varphi}}{t} \,,
\ee
which implies in particular that $\sqrt{-g} \, g^{\mu\nu}$ is
independent of $t$.

Such a scaling solution has a simple interpretation in the limit
where the extra dimensions are large enough to justify a
description in terms of an appropriate low-energy 4D effective
theory \cite{ABPQ}. In the classical limit this theory contains two
massless modes, corresponding to the 4D metric and one combination
of the dilaton, $\phi$, and radius, $r$, of the extra dimensions
(for which $r^2 e^\phi$ is fixed) which parameterizes a flat
direction of the 4D scalar potential. The above scaling solution
describes a time-dependent scaling along this flat direction with
a fixed metric in the 4D Einstein frame. (More general choices for
$c$ also rescale the Einstein-frame 4D metric.)

Explicit scaling solutions to the field equations of the effective
4D theory describing these modes are known, many of which are
attractor solutions to which a broad class of initial conditions
are drawn \cite{ABRS2}. The solutions found here show how to
extend those of the effective 4D theory to see the profiles of the
other nonzero KK modes. Because the 4D solutions are attractors
for the 4D field equations, we might also expect that the same may
be true for the higher-dimensional solutions found here.

\subsubsection*{Pure tension branes}

It is known that even for the special case of pure tension branes,
maximally-symmetric solutions to the 6D field equations
only exist when the tensions of the two branes are adjusted
relative to one another \cite{SLED2,GGP,GGPplus}. One might hope
that the above scaling solutions might describe the late time
behaviour of the solutions in the case that the brane tensions are
not adjusted in the appropriate way.

We now show that a subset of the solutions found above can indeed
describe this situation. In order to do so we must identify when
the asymptotic form of the solutions near the branes have the pure
tension form where $a \sim w$ i.e. $\xi \to 0$. Since the $(\eta
t)$ implies $\xi'=(2+c) z'/(6 c)$ there are only two circumstances
for which $\xi' \to 0$ near a brane:
\be \label{PureTensionCases}
    (i) \quad z' \to 0 \,,
    \qquad\hbox{or}\qquad
    (ii) \quad c=-2  \,.
\ee
\subsubsection*{Special case $\bf c=\infty$}

Case $(i)$ of \pref{PureTensionCases} corresponds to the special
case where the geometry near the brane has a conical singularity,
since this always requires $z' \to 0$ in the near-brane limit.
While this is always possible for one brane, in general it is not
possible for both branes since the equation of motion for $z$ is
\be
    z''-3c H_0^2 e^{2y-2z/c}=0,
\ee
and so $|z''|>0$. There is however, one special case for which
this can be achieved. Redefining $H_0=|c|^{-1}\hat{H}_0$ and
taking the limit $c \to \infty$ we recognize that the originally
coupled system for $y,z$ decouples,
\be
    y''+\frac{4g^2}{\kappa^2} e^{2y}-4 \hat{H}_0^2 e^{2y} =z'' = 0.
\ee
For which the conical solution is
\be
    e^{-y} = \lambda_1^{-1} \sqrt{\left( \frac{4g^2}{\kappa^2}-4 \hat{H}_0^2 \right)}
    \cosh[\lambda_1(\eta-\eta_2)], \quad
    z=z_0.
\ee
As before we can choose $z_0=0$ without loss of generality.
Although the metric, eq.~\pref{MetricScalingAnsatz}, appears to be
singular in this limit, this is only a consequence of an
inconvenient choice for the time coordinate. If we instead convert
to `proper' time, $\tau$, defined by $\exd \tau = t^{c/2} \exd t$,
then $\tau \propto t^{1+c/2}$ and the metric of
eq.~\pref{MetricScalingAnsatz} has a smooth large-$c$ limit:
\be \label{MetricScalingAnsatz1}
    \exd s^2 =  - e^{2w(\eta)}
    \exd \tau^2 + \tau^{2} \Bigl[ e^{2a(\eta)} \, \delta_{ij} \, \exd x^i \exd x^j
    +  e^{2v(\eta)}  \exd \eta^2 +
    e^{2b(\eta)} \exd \theta^2 \Bigr]  \,.
\ee
We now convert back to `conformal' time, $t$, using $\exd
\tau \propto \tau \,\exd t$, then $\tau \sim e^{\hat{H}_0 t/2}$, and
\be \label{MetricScalingAnsatz2}
    \exd s^2 =  e^{\hat H_0 t} \Bigl[ - e^{2 w(\eta)}
    \exd t^2 +  e^{2 a(\eta)} \, \delta_{ij} \, \exd x^i \exd x^j
    +  e^{2v(\eta)}  \exd \eta^2 +
    e^{2b(\eta)} \exd \theta^2 \Bigr]  \,,
\ee
As $\eta \rightarrow \pm \infty$, $e^{ b(\eta)-v(\eta)} \sim (\kappa q)^{-1
}\sqrt{\frac{4g^2}{\kappa^2}-4 \hat{H}_0^2}e^{\pm \lambda_1(\eta_1-\eta_2)}$. Converting to proper
radius $d \rho= e^{ v(\eta)} d\eta$ and identifying with the
conical deficit form $\d\rho^2 +(1-\delta)^2 \rho^2 \d \theta^2$
we infer the deficit angles
\be
    \delta_{\pm} = 1- \lambda_1 \frac{\sqrt{\frac{4g^2}{\kappa^2}
    -4 \hat{H}_0^2}}{\kappa q}e^{\pm \lambda_1(\eta_1-\eta_2)}
\ee
It is clear that by choosing any two of $\lambda_1$, $\eta_1$,
$\eta_2$ and $\hat{H}_0$ appropriately, we can match this solution
onto two conical branes of arbitrary tensions.

Since in conformal time, all the `scale
factors' are exponential, this remains true in 4D Einstein frame.
Such a solution arises when the effective 4D equation of state is
$w=-1/3$, precisely at the transition from acceleration to
deceleration. By itself this behaviour cannot be responsible for
late time dark energy in the form of quintessence.

\subsubsection*{Special case $\bf c=-2$}

Option ($ii$) of \pref{PureTensionCases} makes the choice $c = -2$,
since in this case the $(t\eta)$ Einstein equation implies the
strong statement that $\xi' = \frac14(w - a) = 0$ everywhere
throughout the bulk. For this special case the 6d geometry is
everywhere maximally symmetric in the noncompact 4 dimensions,
taking the form
\be
    \exd s^2 =   e^{2w(\eta)} \d s_{dS^4}^2 +  e^{2v(\eta)}  \exd \eta^2 +
    e^{2b(\eta)} \exd \theta^2  \,,
\ee
where $dS^4$ is 4D de Sitter metric. These are just the de Sitter
geometries considered in \cite{6DdSSUSY}. Notice that there are no
de Sitter solutions where both branes are conical.

\subsection{Asymptotic forms}

We now re-examine the near-brane behaviour of these scaling
solutions in order to connect their properties to those of the
source branes. If we repeat the analysis of asymptotic forms given
earlier for static solutions for the scaling solutions, with
near-brane asymptotic form assumed to be given by
eq.~\pref{MetricScalingAnsatz}, we find two changes relative to
the static case. The simplest change is simply that the assumed
time dependence implies that the coefficients $c_w$, $c_a$ {\it
etc.} now depend explicitly on $t$, with
\be
    c^2_w(t) \propto c^2_a(t) \propto t^c \,, \quad
    c_\theta^2(t) \propto  t^{2+c} \quad
    \hbox{and} \quad
    c^2_\phi(t) \propto t^{-2-c} \,.
\ee

The second change is to do with the relationship amongst the
powers $\alpha$, $\beta$, $\gamma$, $\omega$ and $p$ which is
dictated by the bulk equations. For instance, in the static case
these equations required the powers to be related to one another
by the Kasner-like conditions $\omega + 3\,\alpha + \beta =
\omega^2 + 3\, \alpha^2 + \beta^2 + p^2 = 1$. These conditions
also apply for the scaling solutions, unchanged by $c$ and $H_0$
because the relevant terms in the field equations are subdominant
in powers of $\rho$. However, in the scaling case there is also a
{\it new} constraint, eq.~\pref{tetaConstraint}, coming from the
$(t\eta)$ Einstein equation, which implies the following for the
asymptotic powers:
\be \label{newtetaConstraint}
    (2\,c+1) \omega + 3\,\alpha + (2+c) \, p = 0 \,.
\ee

\subsection{Generalised scaling solutions}

The previous scaling solutions admit a straightforward
generalisation by allowing for the two compact dimensions to scale
differently with time. The new $\ansatz$ is
\be \label{MetricScalingAnsatz3}
    \exd s^2 = (H_0 t)^c  \Bigl[ - e^{2w(\eta)}
    \exd t^2 + e^{2a(\eta)} \, \delta_{ij} \, \exd x^i \exd x^j
    \Bigr] + (H_0 t)^{2+c} \Bigl[ e^{2v(\eta)}  \exd \eta^2 +
    (H_0 t)^{2s} e^{2b(\eta)} \exd \theta^2 \Bigr]  \,,
\ee
with
\be
    A_\theta = (H_0 t)^ s a_\theta(\eta) \qquad \hbox{and} \qquad
    e^\phi = \frac{e^{\varphi(\eta)}}{(H_0 t)^{2+c}}  \,.
\ee
The technique for demonstrating that this is a solution follows as
before. First we can show that all the powers of $t$ drop out and
the equations reduce to a system for the radial profiles alone. As
before it is helpful to perform the redefinition
(\ref{redefinition}) to the $x,y,z,\xi$ variables (we shall choose
$N=1$). In this case the $(\eta t)$ constraint implies
\be
    \xi'=\frac{1}{6(2c+s)} \left( 2e^{-2x} \kappa^2 s a_{\theta}
    a_{\theta}'  +2 s x'+(4+2c+s)z' \right).
\ee
This can be solved for $\xi$ at least formally. In this case, we
do not find a simple action for the system, but the remaining
equations of motion can be expressed as
\ba
    && a_{\theta}''-2a_{\theta}' x' -(1+2c)s H_0^2 e^{2y+z-6 \xi}
    a_{\theta}=0 \nn \\
    && x'' -e^{-6 \xi-2x} \left( e^{2x+2y+z} H_0^2 s(1+2c+s)+e^{2y+z}
    H_0^2 \kappa^2 s^2 a_{\theta}^2-e^{6\xi}\kappa^2 (a_{\theta}')^2
    \right)=0 \nn\\
    && y'' -\frac{1}{\kappa^2} e^{-6\xi-2x+2y} \left(-2 e^{2x}(2 e^{6\xi}g^2
    -e^z H_0^2 \kappa^2 (2+2 c^2+s+s^2+2c(1+s)))+e^{z}H_0^2\kappa^4 s^2
    a_{\theta}^2 \right) =0  \nn\\
    && z'' -\frac{1}{2}e^{-6 \xi-2x+2y+z}H_0^2  \left( e^{2x}(6c-s(s-3))
    -\kappa^2 s^2 a_{\theta}^2\right)=0,
\ea
along with the second constraint
\ba
    &3\left(2c+s\right)^2 \left[ (x')^2+(z')^2-(y')^2+e^{-2x}
    \kappa^2 (a_{\theta}')^2-\frac{4g^2}{\kappa^2} e^{2y} \right. \nn \\
    & \qquad  +e^{2y+z-6\xi}H_0^2 \left(2(2+2c^2+(2c+s)(1+s))
    +e^{-2x}s^2 \kappa^2 a_{\theta}^2\right) \Bigr] \nn \\
    &+\Big[ (4+2c+s)z'+2sx'+e^{-2x} s\kappa \left(a_\theta ^2\right)'
    \Big]^2=0.
\ea
In this case it is necessary to resort to numerics to make further
progress, nevertheless we have demonstrated that such a class of
solutions exists.

\subsubsection*{Scaling symmetry}

True to its name, the generalized scaling solution admits a
scaling symmetry such that under $t \to \lambda t $, $\vec{x} \to
\lambda \vec{x}$ and $\theta \to \lambda^{-s} \theta$, the metric
transforms as $g_{\ssM\ssN} \to \lambda^{2+c} g_{\ssM\ssN}$.
Infinitesimally this corresponds to the existence of a vector V
\be
    V=t \frac{\partial}{\partial t}+x_1 \frac{\partial}{\partial
    x_1}+x_2 \frac{\partial}{\partial x_2}+x_3 \frac{\partial}{\partial
    x_3}-s \theta \frac{\partial}{\partial \theta},
\ee
which satisfies the homethetic Killing vector condition \cite{exactsolutionsbook}
\be
    {\mathcal L}_V g_{\ssM\ssN}=-(D_\ssM V_\ssN+D_\ssN V_\ssM)
    = -a \, g_{\ssM\ssN},
\ee
for constant $a=2+c$. Spacetimes admitting such a vector are also
known as self-similar. This vector is only globally well defined
for $s=0$ since for nonzero $s$ the interval $\theta \in [0,2\pi]$
is not mapped onto itself. The dilaton and gauge field transform
as eigenfunctions of this Killing vector
\ba
    {\mathcal L}_V e^{\phi} &=&-(2+c) e^{\phi}, \nn \\
    {\mathcal L}_V A&=&-(2+c) A.
\ea

\section{Noncompact conical solutions}

In this section we shall present some special time-dependent
solutions that describe conical branes. Unlike the previous
solutions, they describe a single brane in an uncompactified
space. Nevertheless it is reasonable to expect that they could
describe the near brane geometry of a more general time-dependent
two-brane compact solution which is not described by the metrics
already given.

Our $\ansatz$ for the metric is motivated by the form of the
scaling solutions:
\be
    \exd s^2=-\tilde{a}^2(t)\tilde{W}^8(t) \tilde{N}^2(t) \exd t^2 + \tilde{W}^2(t)
    \left( e^{\tilde{\xi}(t)} \exd\vec{x}^2 + e^{-3 \tilde{\xi}(t)} \exd r^2 \right)+
    r^2 \tilde{a}^2(t) \exd\theta^2,
\ee
along with $\phi=\tilde{\varphi}(t)$ and gauge potential
$A_{\theta} =r^2 Q(t)$. Satisfying the $(r,t)$ Einstein equation
requires $Q(t)=Q=$ constant and $\tilde{\xi} = (3\tilde{y} -
2\tilde{z})/12 + \tilde{\xi}_0$. This is equivalent to saying
$\tilde{W}^2 e^{-3\tilde{\xi}} = \tilde{a}^2 e^{-3\tilde{\xi}_0}$,
and so we see that assuming we have normalized $\theta$ to lie on
the interval $[0,2\pi]$, then the metric describes a conical brane
with deficit angle $2\pi(1-e^{3\tilde{\xi}_0})$. It is useful to
make the change of variables
\ba
    && \ln \tilde{a} = (\tilde{x}-\tilde{y}+\tilde{z})/4 \\
    && \ln \tilde{W} = (2 \tilde{x}+\tilde{y})/8 \\
    && \tilde{\varphi} = -(\tilde{x}+\tilde{y}+\tilde{z})/2.
\ea
As usual the remaining equations of motion and constraint can be
derived from the action
\be
    S=\int \d t \left[ \tilde{N}^{-1}\left( 6 \dot{\tilde{x}}^2
    -3 \dot{\tilde{y}}^2-2 \dot{\tilde{z}}^2 \right)+\tilde{N}
    \left( 24 e^{2 \tilde{x}}\frac{g^2}{\kappa^2}+24 e^{2\tilde{x}+
    2\tilde{y}}\kappa^2 q^2 \right) \right]
\ee
This action looks qualitatively similar to the earlier actions,
except for the signs of the kinetic terms.

\section{Brane and bulk dynamics}

In this section we relate the asymptotic form of the above
solutions to brane properties, and use the result to argue that
the solutions capture the late-time
evolution of a pair of brane sources with nontrivial equations of state.

\subsection{Matching to brane properties}

It is possible to make a general statement of how brane properties
dictate the asymptotic forms of the bulk fields in the near-brane
limit, at least for branes for which gravity contributes
negligibly to the total brane stress energy. This section makes
this statement explicit for the 6D case, following arguments
presented in refs.~\cite{NavSant,GGPplus}.

\subsubsection*{A charged aside}

Before doing so, it is worth briefly pausing to develop some
intuition from the analogous problem in electromagnetism. Consider
for this purpose the electrostatic potential, $\varphi(\bfr)$,
generated by a collection of point charges, $Q_i$, situated at
various positions, $\bfr_i(t)$, within 3 spatial dimensions. In
this case we know that some features of the resulting field are
governed purely by the properties of individual charges, while
others depend on the overall configuration of all of the charges.

Typically the field very near the source charges depends purely on
the properties of the nearby source, with the asymptotic behaviour
having Coulomb form, $\varphi(\bfr) \to Q_i/|\bfr - \bfr_i|$, as
$|\bfr - \bfr_i| \to 0$. This form is ultimately dictated by
Gauss' Law, which constrains the local electric field, $\bfE = -
\nabla \varphi$, in terms of the local charge distribution. On the
other hand, whether or not a given charge configuration is static
--- {\it i.e.} whether $\dot \bfr_i = 0$ --- is a type of question
which cannot be purely determined using only near-charge
properties, since it requires knowledge of the global positioning
of {\it all} of the charges.

This same kind of distinction arises also in the case of interest
here: the gravitational fields sourced by a collection of branes.
Again some features of the bulk geometry near the brane are
completely dictated by the physical properties of the brane, while
others --- most notably the time-dependence of the geometry ---
depend on the complete configuration of branes which are present.
The next few sections identify which of the properties of the bulk
fields are which.

\subsubsection*{Thick branes and effective currents}

In this
section we use the bulk field equations to show that local brane
properties determine the near-brane form for radial derivatives (in the Gaussian-normal coordinates of \S\ref{asymptoticforms}) of
the dilaton and gauge potential, $\partial_\rho\phi$ and $F_{\rho
a}$, as well as various combinations of the metric $g_{ab}$ and
its extrinsic curvature, $K_{ab} = \frac12 \, \partial_\rho
{g}_{ab}$. Other quantities depend on the properties of {\it all}
of the branes which source the solution, and so cannot be inferred
purely from the local properties of nearby branes.

To establish these points in the Appendix we review the arguments
of ref.~\cite{NavSant} (see also \cite{GGPplus,deRham:2005ci}) and imagine
regarding the brane source to be a `thick' brane, which physically
extends over a small proper distance, $0 < \rho < \ell$. Within
this region we understand that the microphysical brane structure
modifies the bulk equations, eqs.~\pref{fieldequations}, to
include new sources which are present only for $\rho < \ell$ and
whose presence acts to smooth out the interior geometry at $\rho =
0$. Use of the field equations allows one to relate some of the
near-brane properties of the external bulk fields to particular
averages of these sources over the thick branes.

If in particular these new sources depend only on the radial
coordinate, involve only weak gravitational fields and the
external bulk fields satisfy the asymptotic near-brane power-law
behaviour of eqs.~\pref{asymptoticfields}, then the constants in
this asymptotic form are related to the brane's energy density,
$\varepsilon=-{t^t}_t$, pressures in the three noncompact and
$\theta$ directions, $p_i={t^i}_i$ and
$p_\theta={t^\theta}_\theta$, dilaton `charge', $\sigma$, and
Maxwell current, $j^\theta$, by simple expressions:
\bea
    \kappa^2 \varepsilon &\approx& 2\pi \, \Bigl[1-c_\theta
    (3 \, \alpha + \beta ) (H_1\ell)^{\beta - 1}  \Bigr]
    = 2\pi \, \Bigl[1-c_\theta  (1 - \omega )(H_1\ell)^{\beta - 1}
    \Bigr] \nn\\
    \kappa^2 p_i &\approx& 2\pi \,\Bigl[
    c_\theta  ( \omega + 2\, \alpha + \beta )(H_1\ell)^{\beta
    - 1} - 1\Bigr] \, {\delta^i}_j
    = 2\pi \,\Bigl[ c_\theta  ( 1-\alpha ) (H_1\ell)^{\beta
    - 1}- 1\Bigr] \, {\delta^i}_j \nn\\
    \kappa^2 {t^\theta}_\theta &\approx& 2\pi \,c_\theta
    ( \omega + 3\, \alpha )(H_1\ell)^{\beta - 1}
    = 2\pi \,c_\theta  ( 1 - \beta ) (H_1\ell)^{\beta
    - 1}\nn\\
    \kappa^2 \sigma &=& 2 \pi \, c_\theta p \,(H_1\ell)^{\beta
    - 1} \nn\\
    \hbox{and} \qquad j^\theta &=& 2 \pi \, c_\theta c_f
    (H_1\ell)^{p+\beta-2}
    \,.
\eea
These expressions, whose detailed derivation is given in the
Appendix, identify which features of the near-brane bulk solutions
are governed purely by the local properties of the brane,
reproducing standard results in the case of a conical singularity
with defect angle $2\pi\delta$ in the special case $\alpha =
\omega = p = 0$, $\beta = 1$ and $c_\theta = 1-\delta$. All of the
remaining near-brane bulk-field properties
--- including in particular the time-dependence of the 4D metric
--- cannot be similarly determined purely from local information,
and so depend in detail on the properties of both of the source
branes.

Notice that the above relations impose relations amongst the
stress-energy components ${t^a}_b$ once they are combined with the
bulk field equations. In particular, the relations $\omega + 3\,
\alpha + \beta = 1$ and $\omega^2 + 3\, \alpha^2 + \beta^2 + p^2 =
1$ imply the components of the stress energy tensor must satisfy
\bea
    {t^a}_a = {t^t}_t + \sum_i {t^i}_i + {t^\theta}_\theta
    &=& \frac{8\pi}{\kappa^2} \Bigl[ C_\theta - 1 \Bigr] \nn\\
    ({t^t}_t)^2 + \sum_i ({t^i}_i)^2 + ({t^\theta}_\theta)^2 + \sigma^2
    &=& \frac{8\pi^2}{\kappa^4} \Big[ 2(C_\theta - 1)^2 + C_\theta (1-\beta) \Bigr]
    \,,
\eea
where $C_\theta = c_\theta (H\ell)^{\beta-1}$. For the simplest
case of time-dependent scaling solutions discussed in
\S\ref{constraintssolutions}, these must be supplemented by the
additional relation following from eq.~\pref{newtetaConstraint}:
\be
    (2\,c+1) {t^t}_t + \sum_i {t^i}_i - (2 + c) \, \sigma =
    \frac{4\pi}{\kappa^2} (C_\theta - 1) (c + 2) \,.
\ee
This equation is equivalent to the conservation of stress-energy
on the brane. In general, in the presence of a dilaton coupling
and current, the brane stress energy is not independently
conserved but can couple with the bulk fields.

\subsubsection*{Stress-energies for the scaling solutions}

We are now in a position to calculate the brane stress-energies
that give rise to a specific scaling solution. For simplicity we
shall consider the solutions of \S\ref{constraintssolutions}.
Since the near-brane limit corresponds to $\eta \rightarrow \pm
\infty$, it is sufficient to consider the asymptotic form of the
scaling solutions, for which the metric components behave as
exponentials. After a simple coordinate transformation, we can put
the metric in the Kasner form
\be \label{scalingasymptoticformeqn}
    \d s^2 = (H_0 t)^c \Bigl[ -f_t^{\pm} (H_1r)^{2\omega_{\pm}}\d t^2
    +f_{a}^{\pm }(H_1 r)^{2 \alpha_{\pm}} \delta_{ij}\d x^i \d x^j
    \Bigr]
    +(H_0 t)^{2+c} \left[ \d r^2+f_{\theta}^{\pm} (H_1 r)^{2\beta_\pm-2}
    r^2 \d \theta^2 \right] \, .
\ee
Here the constants $f_t^{\pm}$, $f_a^{\pm}$ and $f_{\theta}^{\pm}$
are determined by integration of the equations of motion. The
Kasner exponents are given by formulae similar to the static
solutions
\ba
    && \alpha_\pm = \frac{\pm \lambda^{\pm}_2 - \lambda_1 }{\pm 5 \lambda^{\pm}_2 - \lambda_1
    \mp 2 \lambda^{\pm}_3} \,,  \quad \quad \quad \omega_\pm =
    \frac{\pm \lambda^{\pm}_2 - \lambda_1 }{\pm 5 \lambda^{\pm}_2 - \lambda_1
    \mp 2 \lambda^{\pm}_3}\, , \\
    && \beta_\pm = \frac{\pm \lambda^{\pm}_2 + 3 \lambda_1 \mp 2\lambda^{\pm}_3}{\pm 5
    \lambda^{\pm}_2 - \lambda_1 \mp 2 \lambda^{\pm}_3} \quad
    \hbox{and} \quad
    p_\pm = -\frac{2( \pm \lambda^{\pm}_2 - \lambda_1 \mp 2 \lambda^{\pm}_3)}{\pm 5
    \lambda^{\pm}_2 - \lambda_1
    \mp 2 \lambda_3}  \,,
\ea
where the $\lambda$'s are defined in \S\ref{constraintssolutions}.
To directly compare with the boundary conditions, we must put the
metric in GN coordinates. This can be achieved approximately with
the transformation $r = (H_0t)^{-1-c/2}\rho$, at the cost of
introducing new $O(\rho^2)$ contributions to the $\exd t^2$ and $\d t \d \rho$ components of the metric. These new contributions are negligible
compared to those listed in eq.~\pref{scalingasymptoticformeqn} in
the $\rho \rightarrow 0$ limit provided only that $\omega < 1$,
which is in practice not a restrictive assumption. The result is a
metric which takes the form
\ba
    \d s^2 &=&  -f_t ^{\pm}(H_0 t)^{c-\omega_{\pm}(2+c)} (H_1\rho)^{2\omega_{\pm}} \d t^2
    +f_{a}^{\pm }(H_0 t)^{c-\alpha_{\pm}(2+c)}
    (H_1\rho)^{2 \alpha_{\pm}} \delta_{ij}\d x^i \d x^j  \nonumber \\
    && \qquad\qquad + \d \rho^2 + f_{\theta}^{\pm} (H_0 t)^{(1-\beta_{\pm})(2+c)}
    (H_1\rho)^{2\beta_{\pm}} \d \theta^2  \, ,
\ea
and the dilaton and form-field are
\ba
    e^{\phi}&=& f^\pm_{\phi} (H_0 t)^{-(2+c)(1+p_{\pm}/2)}
    (H_1\rho)^{p_{\pm}} \, ,\\
    F^{\rho \theta}&=& f^\pm_f (H_0 t)^{-(2+c)(1+p_{\pm}/2)}
    (H_1\rho)^{p_{\pm}-1}
    \,.
\ea
We note that it is important to transform to GN coordinates in
order to make contact with the time-dependence of the brane
geometry. In particular, whilst the coordinate position of the
edge of the brane at $\rho = \ell$ is constant in GN coordinates,
the $t$-dependence of the change of variables between $r$ and
$\rho$ implies that it is instead located along a curve ${\cal
C}(r,t) = \ell$ in another coordinate system.

Finally we can read off the time-dependent brane stress-energies,
dilaton coupling and current,%
\ba
\label{finalresult}
    \kappa^2 \varepsilon^{\pm} &=&
    2\pi \left[ 1 - (1-\omega_{\pm}) f^\pm_{\theta}(H_0 t)^{(1-\beta_{\pm})(1+c/2)}
     (H_1\ell)^{\beta_{\pm}-1} \right] \nonumber \, ,\\
    \kappa^2 p_i^{\pm} &=& -2\pi \left[ 1 - (1-\alpha_{\pm}) f^\pm_{\theta}
    (H_0 t)^{(1-\beta_{\pm})(1+c/2)}
    (H_1\ell)^{\beta_{\pm}-1} \right] \nonumber \, , \\
    \kappa^2 p_{\theta}^\pm &=& 2\pi (1-\beta_{\pm}) f^\pm_{\theta}
    (H_0 t)^{(1-\beta_{\pm})(1+c/2)}
    (H_1\ell)^{\beta_{\pm}-1}\nonumber \, ,\\
    \kappa^2 \sigma_{\pm} &=&  2\pi p_{\pm} f^\pm_{\theta}
    (H_0 t)^{(1-\beta_{\pm})(1+c/2)}
    (H_1\ell)^{\beta_{\pm}-1}\nonumber \, ,\\
    j^{\theta}_{\pm} &=&  2\pi f^\pm_{\theta} f^\pm_f
    (H_0 t)^{-(3+p_{\pm}-\beta_{\pm})(1+c/2)}
    (H_1\ell)^{\beta_{\pm}+p_{\pm}} \,.
\ea
In the special
case of a conical singularity we have $\beta = 1$ and so $\alpha =
\omega = p = 0$, giving:
\be
    \kappa^2 \varepsilon = - \kappa^2 p_i
    = 2\pi \left[ 1 -  f_{\theta} \right] \,, \quad
    \kappa^2 p_{\theta} = \kappa^2 \sigma =  0 \,, \quad
    j^{\theta} =  2\pi f_{\theta} f_f
    (H_0 t)^{-(2+c)}
    (H_1\ell) \,.
\ee
The validity of the expressions (\ref{finalresult}) rests on the
weak-field assumption used in deriving the boundary conditions in
the Appendix. Taking them literally for all time we would infer
that the brane energy becomes negative at some point in the past
or the future, something which is clearly unphysical. This arises
because the effective deficit angle becomes negative. It seems
likely that this is an artifact of our prescription for the
boundary conditions, or that the bulk solutions are no longer
valid in this regime. To resolve this it will be necessary to
build smooth, defect-like models of the branes and analyze the
relationship between the boundary conditions and the bulk
solutions. We feel that nevertheless, the scaling solutions we
have found here will at least represent a consistent description
of the bulk in one or more asymptotic regimes.

\subsubsection*{Delta-function sources and conical singularities}

It is a common practice to represent the brane sources in terms of
a point-like delta-function in the two extra dimensions, with an
action of the form
\be
    S_b = - \int \exd^4x \, \sqrt{-\gamma} \; f(\phi)
    = - \int \exd^6x \, \sqrt{-\gamma} \; f(\phi) \, \delta^2(x) \,,
\ee
where $\gamma_{\mu\nu} = g_{\ssM\ssN} \partial_\mu x^\ssM
\partial_\nu x^\ssN$ denotes the induced metric on the brane. In this
case direct variation of the brane action with respect to bulk
fields would lead to
\be \label{deltafnform}
    T^{\ssM\ssN}_{(b)} = t^{\mu\nu}_{(b)}\, \delta^\ssM_\mu \delta^\ssN_\nu
    \, \frac{\delta^2(x)}{\sqrt{g_2}} \,, \qquad
    J^\ssN_{(b)} = j^\nu_{(b)} \, \delta^\ssN_\nu \, \frac{\delta^2(x)}{\sqrt{g_2}}
     \qquad \hbox{and} \qquad
    P_{(b)} = \sigma_{(b)} \, \frac{\delta^2(x)}{\sqrt{g_2}}
     \,,
\ee
with
\be
    t^{\mu\nu}_{(b)} = \frac{2}{\sqrt{-\gamma}} \, \frac{\delta S_b}{\delta
    \gamma_{\mu\nu}} = - f(\phi) \, \gamma^{\mu\nu}
    \,, \qquad
    \sigma_{(b)} = - \frac{1}{\sqrt{-\gamma}} \, \frac{\delta S_b}{\delta \phi}
    = f'(\phi) \,,
\ee
and $j^\nu_{(b)} = -\delta S_b/\delta A_\nu  = 0$.

However, it is important to realize that such an assumption
requires that the bulk fields remain regular at the brane position,
$\rho \to 0$, so that it makes sense to evaluate the bulk fields
there. The above expressions show that this is {\it not} generic
for branes having co-dimension $\ge 2$, since it requires the
powers $\alpha$, $\beta$, $\omega$ and $p$ all to be non-negative.
If the brane couples to the Maxwell field then the condition on
$p$ strengthens to $p \ge 1$, which is consistent with the bulk
field equation $\alpha^2 + \beta^2 + \omega^2 + p^2 = 1$ only if
$p = 1$ and $\alpha = \beta = \omega = 0$. Otherwise the
representation of the brane in terms of a $\delta$-function source
can be overly restrictive. More generally one can instead describe
the low-energy bulk dynamics by excising the brane positions from
the bulk geometry, and describing the brane sources in terms of
boundary dynamics on the resulting co-dimension one boundary of
the excised region \cite{BoundaryAction}.

\subsection{Parameter counting}

\S\ref{static} provides an 8-parameter family of static
solutions, with independent parameters $\lambda_2$, $\lambda_3$,
$\lambda_4$, $q$, $\eta_1$, $\eta_2$, $z_0$ and $\xi_0$.
$\lambda_1$ can be fixed by means of the constraint equation
$\lambda_2^2=\lambda_1^2+\lambda_3^2+12 \,\lambda_4^2$. Of these,
two (say $\eta_2$ and $\xi_0$) can be removed by appropriately
choosing units in the noncompact four dimensions, since this can
be used to set $a$ and $w$ to any convenient value at any one
position. (Notice that $b(\eta_0)$ cannot be similarly set to
vanish in this way without changing the periodicity condition
$\theta \simeq \theta + 2\pi$. We also do not use the classical
scale invariance of the bulk field equations to remove $z_0$ by
allowing $\phi$ to be set to any desired value at any one point,
$\eta = \eta_0$. We do not do so because this scale invariance may
be broken by the brane-bulk couplings, and so may not be
consistent with the boundary conditions given in the previous
section.) This leaves a total of $8-2=6$ physical parameters in
the static solutions presented above.

A similar counting applies to the generalised scaling solutions,
whose time dependence introduces three more parameters: $c$, $s$
and $H_0$.\footnote{We note that $H_1$ is not an independent
parameter.} Integrating the remaining equations to obtain the
field profiles as functions of $\eta$ may be expected to introduce
only 5 more constants (rather than 6, due to the additional
Bianchi-identity related constraint relative to the static case)
into the general scaling solution as well, leading us to expect
there to be a total of $5 + 3 = 8$ parameters in these solutions.

We may now count the number of parameters which should be expected
of a general bulk configuration which is sourced by two branes.
For the symmetries of interest each brane is characterized by the
5 quantities $\varepsilon$, $p_i$, $p_\theta$, $j^\theta$ and
$\sigma$. With two source branes there are 10 such parameters in
total. To this should also be added the integer which measures the
total magnetic flux through the extra dimensions, so long as we
also include the topological constraint \cite{SLED2} which relates
this integer to the magnetic currents on each brane. This leaves
us with a total of 10 independent parameters describing the
physics which sources the bulk fields, one of which is quantized
to be an integer. This represents 2 more parameters than our
general time-dependent solutions have available to accomodate. We
believe this is because the solutions we consider are not singular
enough to contain direct couplings between the branes and the
Maxwell fields, and so $j^\theta = 0$ on both branes.

Scaling solutions are often attractor solutions towards which
general time-dependent configurations tend after the passage of
any initial transients. If this is also so for the 6D supergravity
field equations, we would be led to the following attractive
picture. It has long been known \cite{SLED2,GGP,GGPplus} that
static (and maximally-symmetric, but curved \cite{6DdSSUSY})
solutions can only exist if the properties of the two source
branes are appropriately adjusted relative to one another. But it
has been unknown what happens to the bulk geometry in the generic
case where such adjustments are not made, although it has been
suspected that these would produce time-dependent bulk
configurations. Based on the above considerations, in the generic
case we expect that the bulk is indeed time-dependent, and in
particular this time-dependence approaches one of the scaling
solutions given here (once transients pass) at late times.

\section{Conclusions}

Six dimensional supergravity provides a fascinating laboratory for
investigating the issues which underly the cosmological constant
problem, largely because six dimensions is both simple enough to
allow the construction of explicit solutions, yet rich enough to
exhibit an interesting variety of properties. In particular, it
provides the simplest setting within which a collection of
positive-tension branes can combine to produce vanishing 4D
curvature. This makes it a very fruitful arena in which to explore
how natural are the choices which must be made in order to ensure
acceptably flat 4D worlds.

Our main result in this paper is to provide a new class of static
and time-dependent solutions to the full field equations of gauged
chiral 6D supergravity. For both classes of solutions there are 4
warped, noncompact dimensions (which need not be maximally
symmetric), and 2 curved compact dimensions which are taken to be
axially symmetric. The time-dependent solutions are of the
self-similar scaling type, and as such are plausibly attractors
which robustly describe the late-time limit of the time-dependent
evolution generated by fairly generic initial conditions.

The solutions we find diverge at (at most) two positions within
the extra dimensions, which we interpret to be the positions of
two space-filling 3-branes. This interpretation is supported by
examining the asymptotic behaviour near the branes, which has a
relatively simple interpretation in terms of brane stress-energy
and dilaton coupling, at least in the limit that the gravitational
fields involved in the brane structure are very weak.

The picture which emerges of brane-bulk dynamics is this. Much
like for a collection of static electric charges within a compact
space, the fields produced by a generic set of sources is time
dependent. It is nonetheless possible to achieve static solutions
provided the properties of the two branes involved are
appropriately adjusted relative to one another. In the
supersymmetric case the resulting static solutions are also
marginally stable (classically) \cite{Perturbations,PRS}, with the
marginal perturbations being along a flat direction whose presence
is guaranteed by a classical scale invariance of the 6D
supergravity equations. We believe our scaling solutions to be the
natural endpoint for motions which start out along these flat
directions.

What do these results imply for the 6D self-tuning mechanism for
the cosmological constant? It must be emphasized that they do not
at all address the central issue of technical naturalness --- {\it
i.e.} the question of whether the choices required for flat (or
slowly-varying) cosmologies are stable against renormalization.
What they do address is the ancillary issue of initial conditions,
which arises in any theory for which the Dark Energy density is
evolving in time. They do so by identifying how special the
solutions are which can give acceptable 4D cosmologies.

Although such acceptable solutions do exist \cite{ABRS2}, their
existence requires both an acceptably shallow potential {\it and}
logarithmic corrections to this potential (such as can be
generated by Casimir energies in the 2 extra dimensions
\cite{UVSensitivity}). However the slow roll due to potentials
like these coming from a quantum origin can only be relevant if
they are not dominated by the fast motion driven by larger
classical forces, and the motion described by the generic
solutions obtained here would be much too fast to provide an
acceptable 4D cosmology. So any description of the Dark Energy in
terms of 6D dynamics along these lines must presuppose initial
conditions which exclude the generic motion found here, and so
rely on the universe being prepared with branes whose properties
are sufficiently well adjusted relative to one another to give
close to static classical dynamics. One might hope that such
special initial conditions might have their explanation in the
same way as did the initial condition problems of the standard Hot
Big Bang cosmology: that is, in terms of the (possibly
inflationary) dynamics of still-earlier epochs of the universe.

\section*{Acknowledgements}

AJT is supported in part by US Department of Energy Grant
DE-FG02-91ER40671. CB would like to thank the Banff International
Research Station for its kind hospitality while this paper was
being finished. CB and DH acknowledge support from the Natural
Sciences and Engineering Research Council of Canada, with
additional support for CB coming from the Killam Foundation and
McMaster University. DH acknowledges support from les fonds
Qu\'eb\'ecois de la recherche sur la nature et les technologies.
CdR is funded by a grant from the Swiss National Science
Foundation. Research at Perimeter Institute is supported in part
by the Government of Canada through NSERC and by the Province of
Ontario through MEDT.

\appendix

\section{Brane bulk matching}

In this Appendix we summarize the connection between the
near-brane asymptotic behaviour of the bulk solutions and some of
the physical properties of these branes, following the arguments
made in refs.~\cite{GGP,NavSant}. To this end we choose Gaussian-normal coordinates as described in \S\ref{asymptoticforms}.

Our goal is to use the bulk field equations to show that local
brane properties determine the near-brane form of the radial
derivatives of the bulk fields: $\partial_\rho\phi$, $F_{\rho a}$
and $K_{ab} = \frac12 \, \partial_\rho {g}_{ab}$. Other
quantities, such as the time dependence of the near-brane metric,
depend on the properties of global properties involving {\it all}
of the branes which source the solution, and so cannot be inferred
purely from the properties of the brane at $\rho = 0$.

To establish these points we follow refs.~\cite{GGPplus,NavSant,deRham:2005ci}
and imagine regarding the brane source to be a `thick' brane,
which physically extends over a small proper distance, $0 < \rho <
\ell$. Within this region we understand that the microphysical
brane structure modifies the bulk equations,
eqs.~\pref{fieldequations}, to include new sources in such a way
as to ensure that all quantities remain smooth as $\rho \to 0$. We
quantify this smoothness by assuming that the resulting bulk
fields obey the conditions $K_{\mu\nu} = O(\rho^2)$,
$K_{\theta\theta} = 2\rho$, $F^{a\rho} = O(\rho)$ and
$\partial_\rho \phi = O(\rho)$ as $\rho \to 0$. We define the new
sources which are required in this way from the equations
\be
    G_{\ssM\ssN} + \kappa^2  T_{\ssM\ssN}  =
    D_\ssM F^{\ssM\ssN} - J^\ssN =
    \Box \phi - \kappa^2 P =
    0 \,,
\ee
where $G_{\ssM\ssN}$ is the Einstein tensor. The source currents
$J_\ssN = J^{(B)}_\ssN + J^{(b)}_\ssN$ and $P = P^{(B)} + P^{(b)}$
include both the contributions of the bulk fields, `$(B)$', as
inferred from eqs.~\pref{fieldequations} plus new brane
contributions, `$(b)$', which vanish for $\rho > \ell$. The stress
energy receives similar kinds of contributions, $T_{\ssM\ssN} =
T^{(B)}_{\ssM\ssN} + T^{(b)}_{\ssM\ssN}$.

Now comes the main argument. Consider for simplicity a purely
radial source profile within the thick brane, with
\be
    \exd s^2 = g_{\mu\nu}(\rho) \, \exd x^\mu \exd x^\nu +
    g_{mn}(\rho) \, \exd x^m \exd x^n = g_{\mu\nu}(\rho) \,
    \exd x^\mu \exd x^\nu +
    e^{2b(\rho)} \, \exd \theta^2 + \exd\rho^2 \,,
\ee
and so on for the other bulk fields. The point of the definitions
given above is that the field equations allow the sources to be
written as total derivatives, allowing their averages over the
thick-brane volume\footnote{Multipole generalizations can be
similarly defined by appropriately weighting the integrands by
functions of $\theta$.} to give expressions depending only on the
boundary values at $\rho = 0$ and $\rho = \ell$. This is simplest
to see for the dilaton, for which
\bea \label{sigmadefn}
    \kappa^2 \sigma &\equiv&
    \int_{\rho < \ell} \exd^2x \, \sqrt{g_2} \; \kappa^2 P
    = \int_{\rho < \ell} \exd^2x \, \sqrt{g_2}
    \; \Box \phi \nn\\
    &=& \int_0^{2\pi} \exd \theta \, \Bigl[
    \sqrt{g_2} \; s^\ssM \partial_{\ssM} \phi \Bigr]^{\rho = \ell}_{\rho = 0}
    = 2\pi \Bigl[ e^{b} \, \partial_\rho \phi \Bigr]_{\rho = \ell}
    \,,
\eea
where $s^\ssM$ is the outward-pointing unit normal to surfaces of
fixed $\rho$ and the last equality uses the above-mentioned
requirement that $\partial_\rho \phi$ must vanish at $\rho = 0$.
Eq.~\pref{sigmadefn} gives an explicit relation between
$\partial_\rho \phi |_{\rho = \ell}$ and purely local brane
properties.

A similar argument may be made for the current $J^\ssM$. Averaging
over the thick brane volume gives
\bea
    j^a &=& \int_{\rho < \ell} \exd^2x \,
    \sqrt{g_2} \;  J^a
    =  \int_{\rho < \ell} \exd^2x \,
    \sqrt{ g_2} \; D_\ssM F^{\ssM a}  \nn\\
    &=& \int_0^{2\pi} \exd\theta
    \Bigl[ \sqrt{g_2} \, s_\ssM
    F^{\ssM a} \Bigr]^{\rho = \ell}_{\rho = 0} =
    2\pi  \, \Bigl[ e^{b} \;  F^{\rho a} \Bigr]_{\rho = \ell}
    \,,
\eea
with the contribution from $\rho = 0$ again vanishing due to the
condition that $F^{\rho \ssN}$ must vanish there. We see from this
how the near-brane limit of $F^{\rho a}$ relates to local brane
properties.

A similar result for the energy density is more difficult to
define in general, however progress is possible for purely radial
profiles if the gravitational binding energy of the thick brane is
negligible. In this case, averaged stress energy components can be
defined by
\be
    \kappa^2 {t^a}_b =
    \int_{\rho < \ell} \exd^2x \,\sqrt{g_2} \; \kappa^2 {T^a}_b
    = -  \int_{\rho < \ell} \exd^2x \,
    \sqrt{g_2} \; {G^a}_b  \,.
\ee
It is useful at this point to specialize to the particular metric,
$\exd s^2 = -e^{2w(\rho)} \, \exd t^2 + e^{2a(\rho)} \,
\delta_{ij}\, \exd x^i \exd x^j + e^{2b(\rho)} \, \exd\theta^2 +
\exd \rho^2$, for which $\sqrt{g_2} = e^{b}$ and the nonvanishing
components of the Einstein tensor are
\bea
    -{G^t}_t &=& 3 \, a'' + b'' + 6\, (a')^2 + 3 \,a' b' + (b')^2
    \nn\\
    -{G^i}_j &=& \Bigl[ w'' + 2\, a'' + b'' + 2 \, w' a' + 3 \,(a')^2
    +2 \,a' b' + (w')^2 + w' b' + (b')^2 \Bigr] \, {\delta^i}_j
    \nn\\
    -{G^\theta}_\theta &=& w'' + 3\, a'' + 3\, w' a' + (w')^2 + 6\, (a')^2 \,,
\eea
where $'=d/d\rho$. The main point now is this: although the above
integrals depend on the detailed metric profiles, the result
simplifies in the limit of weak gravitational fields for which the
expression may be linearized in the derivatives $w'$, $a'$ and
$\tilde{b}'$, where $e^{2 b}=\rho^2 e^{2\tilde{b}}$. In this case
the integrals may be performed explicitly, leading to the
expressions
\bea
    \kappa^2 {t^t}_t &\approx& 2\pi \, \Bigl[
    e^{b} (3 \, a' + b' ) \Bigr]^{\rho = \ell}_{\rho = 0} \nn\\
    \kappa^2 {t^i}_j &\approx& 2\pi \, \Bigl[
    e^{b} ( w' + 2\, a' + b' ) \Bigr]^{\rho = \ell}_{\rho = 0}
    \, {\delta^i}_j
    \nn\\
    \kappa^2 {t^\theta}_\theta &\approx& 2\pi \,  \Bigl[
    e^{b} ( w' + 3\, a' ) \Bigr]^{\rho = \ell}_{\rho = 0} \,.
\eea

We now specialize the above expressions by evaluating them using
the smooth limit at $\rho = 0$ and the asymptotic form of the bulk
metrics near $\rho = \ell$:
\bea
    \exd s^2 &\sim& - [c_w (H\rho)^{\omega}]^2 \, \exd t^2
    + [c_a (H\rho)^{\alpha}]^2 \, \delta_{ij} \, \exd x^i \exd x^j
    +  \exd \rho^2 + [c_\theta (H\rho)^{\beta-1}]^2 \rho^2 \exd \theta^2  \nn\\
    F^{\rho\theta} &\sim& c_f (H\rho)^\gamma \qquad \hbox{and} \qquad
    e^{\phi} \sim c_\phi (H\rho)^p \,.
\eea
Here $H$ is an arbitrary dimensionful scale, while $\omega$,
$\alpha$, $\beta$, $p$, $c_w$, $c_a$, $c_\theta$, $c_f$ and
$c_\phi$ are constants, for which the bulk field equations imply
the conditions $\omega + 3 \alpha + \beta = \omega^2 + 3\alpha^2 +
\beta^2 + p^2 = 1$ and $\gamma = p - 1$. For this metric the
relevant derivatives are $w'(\ell) = \omega/\ell$, $a'(\ell) =
\alpha/\ell$, $b'(\ell) = \beta/\ell$ and $\phi'(\ell) = p/\ell$,
while the 2D volume element is $\sqrt{g_2} = c_\theta \,
(H\rho)^{\beta}$.

The relevant quantities at $\rho = 0$ are $e^{b} w'|_0 = e^{b} a'
|_0 = 0$ and $e^{b} b' |_0 = 1$, and so up to contributions that
are sub-leading for small $\ell$ we find
\bea
    \kappa^2 {t^t}_t &\approx& 2\pi \, \Bigl[c_\theta
    (3 \, \alpha + \beta ) (H\ell)^{\beta - 1} - 1 \Bigr]
    = 2\pi \, \Bigl[c_\theta  (1 - \omega )(H\ell)^{\beta - 1}
    - 1 \Bigr] \nn\\
    \kappa^2 {t^i}_j &\approx& 2\pi \,\Bigl[
    c_\theta  ( \omega + 2\, \alpha + \beta )(H\ell)^{\beta
    - 1} - 1\Bigr] \, {\delta^i}_j
    = 2\pi \,\Bigl[ c_\theta  ( 1-\alpha ) (H\ell)^{\beta
    - 1}- 1\Bigr] \, {\delta^i}_j \nn\\
    \kappa^2 {t^\theta}_\theta &\approx& 2\pi \,c_\theta
    ( \omega + 3\, \alpha )(H\ell)^{\beta - 1}
    = 2\pi \,c_\theta  ( 1 - \beta ) (H\ell)^{\beta
    - 1}\nn\\
    \kappa^2 \sigma &=& 2 \pi \, c_\theta p \,(H\ell)^{\beta
    - 1} \nn\\
    \hbox{and} \qquad j^\theta &=& 2 \pi \, c_\theta c_f
    (H\ell)^{\gamma+\beta-1}
    \,.
\eea

These are the expressions which are used in the main body of the
text. Notice that in the case of a purely conical singularity
having defect angle $0 < \delta < 1$ in the bulk geometry we have
$\alpha = \omega = p = 0$ and $\beta = 1$ and $c_\theta = 1 -
\delta$, in which case the above expressions simplify to the
following results for the energy density, $\varepsilon$, 3D pressure,
$p_i$, and off-brane pressure, $p_\theta$:
\be
    \kappa^2 \varepsilon = - \kappa^2 {t^t}_t \approx
    2\pi \,\delta \,,\quad
    \kappa^2 p_i = \kappa^2 {t^i}_i \approx -2\pi \,\delta
    \quad \hbox{and} \quad
    \kappa^2 p_\theta = \kappa^2 {t^\theta}_\theta
    \approx  0 \,.
\ee

\end{document}